\documentclass[runningheads]{llncs}
\usepackage{graphicx}
\usepackage{tikz}
\usepackage{subfig}
\tikzset{> = latex} %
\tikzstyle{c} = [draw, shape=circle, inner sep=0pt, text width=2mm, align=center] %

\usepackage{multirow}

\renewcommand{\sc}[1]{\textsc{#1}}

\newcommand{\npc}{\sc{NP}\normalfont{-complete}}

\usepackage{enumerate}
\usepackage{todonotes}
\usepackage{hyperref}
\usepackage[misc]{ifsym}

\usepackage{mathtools}
\newcommand{\assign}{\vcentcolon=}

\newcommand\restr[2]
{{
  \left.\kern-\nulldelimiterspace
  #1 
  \vphantom{\big|}
  \right|_{#2}
}}

\begin{document}
\title{Efficient Implementation of Color Coding Algorithm for Subgraph Isomorphism Problem\thanks{Extended abstract of this paper will appear in the proceedings of the Special Event on Analysis of Experimental Algorithms, SEA\^{}2 2019, Lecture Notes in Computer Science, Springer.}}
\titlerunning{Color Coding Algorithm for Subgraph Isomorphism}
\author{Josef Mal{\' i}k\thanks{Supported by grant 17-20065S of the Czech Science Foundation.} \and
Ond{\v r}ej Such{\' y}\thanks{The author acknowledges the support of the OP VVV MEYS funded project
CZ.02.1.01/0.0/0.0/16\_019/0000765 ``Research Center for Informatics''.}\and
Tom{\' a}{\v s} Valla$^{\star\,\star\,\star}$}
\authorrunning{J. Mal{\' i}k, O. Such{\' y} and T. Valla}
\institute{Department of Theoretical Computer Science, Faculty of Information Technology, Czech Technical University in Prague, Prague, Czech Republic
\email{josef.malik@fit.cvut.cz}\\
\email{ondrej.suchy@fit.cvut.cz}, {\scriptsize ORCID: 0000-0002-7236-8336}\\
\email{tomas.valla@fit.cvut.cz}, {\scriptsize ORCID: 0000-0003-1228-7160}}
\maketitle              %
\begin{abstract}
We consider the subgraph isomorphism problem where, given two graphs~$G$ (source graph) and~$F$ (pattern graph), one is to decide whether there is a (not necessarily induced) subgraph of~$G$ isomorphic to~$F$.
While many practical heuristic algorithms have been developed for the problem, as pointed out by McCreesh et al. [JAIR 2018], for each of them there are rather small instances which they cannot cope. Therefore, developing an alternative approach that could possibly cope with these hard instances would be of interest.

A seminal paper by Alon, Yuster and Zwick [J. ACM 1995] introduced the color coding approach to solve the problem, where the main part is a dynamic programming over color subsets and partial mappings.
  As with many exponential-time dynamic programming algorithms, the memory requirements constitute the main limiting factor for its usage.
  Because these requirements grow exponentially with the treewidth of the pattern graph, all existing implementations based on the color coding principle restrict themselves to specific pattern graphs, e.g., paths or trees.
  In contrast, we provide an efficient implementation of the algorithm significantly reducing its memory requirements so that it can be used for pattern graphs of larger treewidth.
  Moreover, our implementation not only decides the existence of an isomorphic subgraph, but it also enumerates all such subgraphs (or given number of them).
  
  We provide an extensive experimental comparison of our implementation to other available solvers for the problem.

\keywords{Subgraph isomorphism \and
          subgraph enumeration \and 
          color coding \and 
          tree decomposition \and
          treewidth.}
\end{abstract}

\section{Introduction}

Many real-world domains incorporate large and complex networks of interconnected units.
Examples include social networks, the Internet, or biological and chemical systems.
These networks raise interesting questions regarding their structure.
One of those questions asks whether a~given network contains a~particular pattern, which typically represents a specific behaviour of interest \cite{mccc,ri,cit1}.
The problem of locating a particular pattern in the given network can be restated as a problem of locating a subgraph isomorphic to the given pattern graph in the network graph.

Formally, the \textsc{Subgraph Isomorphism (SubIso)} problem is, given two undirected graphs~$G$ and~$F$, to decide whether there is a (not necessarily induced) subgraph of~$G$ isomorphic to~$F$. Or, in other words, whether there is an adjacency-preserving injective mapping from vertices of $F$ to vertices of $G$.
Since we do not require the subgraph to be induced (or the mapping to preserve non-adjacencies), some authors call this variant of the problem \textsc{Subgraph Monomorphism}. %

For many applications it is not enough to just learn that the pattern does occur in the network, but it is necessary to actually obtain the location of an occurrence of the pattern or rather of all occurrences of the pattern \cite{cit2,pcc}.
Because of that, we aim to solve the problem of subgraph enumeration, in which it is required to output all subgraphs of the network graph isomorphic to the pattern graph. In \textsc{Subgraph Enumeration (SubEnum)}, given again two graphs $G$ and $F$, the goal is to enumerate all subgraphs of~$G$ isomorphic to~$F$.
Note, that \sc{SubEnum} is at least as hard as \sc{SubIso}. We call the variants, where the problem is required to be induced \textsc{IndSubIso} and \textsc{IndSubEnum}, respectively.

As \textsc{Clique}, one of the problems on the Karp's original list of 21 NP-complete problems~\cite{Karp72},
 is a special case of \textsc{SubIso}, the problem is NP-complete.
Nevertheless, there are many heuristic algorithms for \textsc{SubEnum}, many of them based on ideas from constraint programming (see \autoref{sec:related}), which give results in reasonable time for most instances. 
However, for each of them there are rather small instances which they find genuinely hard, as pointed out by McCreesh et al.~\cite{McCreeshPST18}. Therefore, developing an alternative approach that could possibly cope with these hard instances would be of interest.

In this paper we focus on the well known randomized color coding approach~\cite{ayz}, which presumably has almost optimal worst case time complexity.
Indeed, its time complexity is $\mathcal{O}\big(n_G^{\sc{tw}(F) + 1}2^{\mathcal{O}(n_F)}\big)$ with memory requirements of $\mathcal{O}\big(n_G^{\sc{tw}(F) + 1}\sc{tw}(F)n_F 2^{n_F}\big)$,
where~$n_G$ and~$n_F$ denote the number of vertices in the network graph~$G$ and the pattern graph~$F$, respectively, and
$\sc{tw}(F)$ is the treewidth of graph~$F$---a measure of tree-likeness (see \autoref{sec:defs} for exact definitions).
Moreover, we presumably cannot avoid the factor exponential in treewidth in the worst case running time, as Marx~\cite{cybtw} presented an ETH\footnote{Exponential Time Hypothesis~\cite{eth}}-based lower bound for \textsc{Partitioned Subgraph Isomorphism} problem.

\begin{proposition}[Marx~\cite{cybtw}]
\label{eth_bound}
If there is a recursively enumerable class~$\mathcal{F}$ of graphs with unbounded treewidth, an algorithm~$\mathcal{A}$, and an arbitrary function~$f$ such that~$\mathcal{A}$
correctly decides every instance of \textsc{Partitioned Subgraph Isomorphism} with the smaller graph~$F$ in~$\mathcal{F}$ in time $f(F)n_{G}^{o({\sc{tw}(F)}/{\log \sc{tw}(F)})}$, then ETH fails.
\end{proposition}

As the memory requirements of the color coding approach grow exponentially with treewidth of the pattern graph, existing implementations for subgraph enumeration based on this principle restrict themselves to paths~\cite{sig} or trees~\cite{pcc}, both having treewidth~1. As the real world applications might employ networks of possibly 
tens to hundreds of thousands of vertices and also pattern graphs with structure more complicated than trees, we need to significantly reduce the memory usage of the algorithm.

Using the principle of inclusion-exclusion, Amini et al.~\cite[Theorem 15]{AminiFS12} suggested a modification of the color coding algorithm, which can decide whether the pattern $F$ occurs in the graph $G$ in expected time $\mathcal{O}\big(n_G^{\sc{tw}(F) + 1}2^{\mathcal{O}(n_F)}\big)$ with memory requirements reduced to $\mathcal{O}\big(n_G^{\sc{tw}(F) + 1} \log n_F)$.\footnote{While the formulation of Theorem 15 in \cite{AminiFS12} might suggest that the algorithm actually outputs a witnessing occurrence, the algorithm merely decides whether the number of occurrences is non-zero (see the proof of the theorem).} While single witnessing occurrence can be found by means of self-reduction (which is complicated in case of randomized algorithm), the inclusion-exclusion nature of the algorithm does not allow to find all occurrences of pattern in the graph, which is our main goal. 

Therefore, our approach rather follows the paradigm of generating only those parts of a dynamic programming table that correspond to subproblems with a positive answer, recently called ``positive instance driven'' approach~\cite{Tamaki17}.  This further prohibits the use of the inclusion-exclusion approach of Amini et al.~\cite{AminiFS12}, since the inclusion-exclusion approach tends to use most of the table and the term $\mathcal{O}\big(n_G^{\sc{tw}(F) + 1}\big)$ is itself prohibitive in the memory requirements for $\sc{tw}(F) \ge 2$.

Because of the time and memory requirements of the algorithm, for practical purposes we restrict ourselves to pattern graphs with at most~$32$ vertices.

Altogether, our main contribution is twofold:
\begin{itemize}
  \item We provide a practical implementation of the color coding algorithm of Alon, Yuster, and Zwick~\cite{ayz} capable of processing large networks and (possibly disconnected) pattern graphs of small, yet not a priory bounded, treewidth.
  \item We supply a routine to extract the occurrences of the subgraphs found from a run of the algorithm.
\end{itemize}

It is important to note that all the modifications only improve the practical memory requirements and running time. The theoretical worst case time and space complexity remain the same as for the original color coding algorithm and the algorithm achieves these, e.g., if the network graph is complete. Also, in such a case, there are $n_G^{\Theta(n_F)}$ occurrences of the pattern graph in the network implying a lower bound on the running time of the enumeration part.

In \autoref{sec:algorithm} we describe our modifications to the algorithm and necessary tools used in the process. Then, in \autoref{sec:experiments}, we benchmark our algorithm on synthetic and realistic data and compare its performance with available existing implementations of algorithms for subgraph isomorphism and discuss the results obtained. \autoref{sec:conclusion} presents future research directions.

\subsection{Related work} \label{sec:related}

There are several algorithms tackling \textsc{SubIso} and its related variants. Some of them only solve the variant of subgraph counting, our main focus is however on 
algorithms actually solving \textsc{SubEnum}. 
Following Carletti et al.~\cite{vf2plus} and Kimmig et al.~\cite{sharedmem}, we categorize the algorithms by the approach they use (see also Kotthoff et al.~\cite{KotthoffMS16} for more detailed description of the algorithms).
Many of the approaches can be used both for induced and non-induced variants of the problem, while some algorithms are applicable only for one of them.

Vast majority of known algorithms for the subgraph enumeration problem is based on the approach of representing the problem as a searching process.
Usually, the state space is modelled as a tree and its nodes represent a state of a partial mapping.
Finding a solution then typically resorts to the usage of DFS in order to find a path of mappings in the state space tree which is compliant with isomorphism requirements. 
The efficiency of those algorithms is largely based on early pruning of unprofitable paths in the state space. 
Indeed, McCreesh et al.~\cite{McCreeshPST18} even measure the efficiency in the number of generated search tree nodes.
The most prominent algorithms based on this idea are Ullmann's algorithm~\cite{ullmann}, VF algorithm and its variants~\cite{vf3,vf2plus,vf,vf2} (the latest VF3~\cite{vf3} only applies to \textsc{IndSubEnum}) and RI algorithm~\cite{ri}.
The differences between these algorithms are based both on employed pruning strategies and on the order in which the vertices of pattern graph are processed (i.e. in the shape of the state space tree).

Another approach is based on constraint programming, in which the problem is modelled as a set of variables (with respective domains)
and constraints restricting simultaneous variable assignments. The solution is an assignment of values to variables in a way such that no constraint remains unsatisfied.
In subgraph isomorphism, variables represent pattern graph vertices, their domain consists of target graph vertices to which they may be mapped and
constraints ensure that the properties of isomorphism remain satisfied.
Also in this approach, a state space of assignments is represented by a search tree, in which non-profitable branches are to be filtered.
Typical algorithms in this category are LAD algorithm~\cite{lad}, Ullmann's bitvector algorithm~\cite{ullmann_csp}, and Glasgow algorithm~\cite{McCreeshP15}.
These algorithms differ in the constraints they use, the way they propagate constraints, and in the way they filter state space tree. 

There are already some implementations based on the color coding paradigm, where the idea is to randomly color the input graph and search only for its subgraphs, isomorphic to the pattern graph, that
are colored in distinct colors (see \autoref{subs:idea} for more detailed description). This approach is used in subgraph counting algorithms, e.g., in ParSE~\cite{parse}, FASCIA~\cite{fascia}, and in~\cite{mccc}, or in algorithms for path enumeration described in~\cite{pcc} or in~\cite{sig}.
Each of these algorithms, after the color coding step, tries to exploit the benefits offered by this technique in its own way; although usually a dynamic programming sees its use.
Counting algorithms as ParSE and FASCIA make use of specifically partitioned pattern graphs, which allow to use combinatorial computation.
Weighted path enumeration algorithms \cite{pcc,sig} describe a dynamic programming approach and try to optimize it in various ways.
However, to the best of our knowledge there is no color coding algorithm capable of enumerating patterns of treewidth larger than~1.

Our aim is to make step towards competitive implementation of color coding based algorithm for \textsc{SubEnum}, in order to see, where this approach can be potentially beneficial against the existing algorithms. To this end, we extend the comparisons of \textsc{SubEnum} algorithms~\cite{compar,KotthoffMS16,McCreeshPST18} to color coding based algorithms, including the one proposed in this paper.

\subsection{Basic definitions}\label{sec:defs}
All graphs in this paper are undirected and simple.
For a graph~$G$ we denote~$V(G)$ its vertex set,~$n_G$ the size of this set,~$E(G)$ its edge set, and~$m_G$ the size of its edge set.

As already said, we use the color coding algorithm. The algorithm is based on a dynamic programming on a nice tree decomposition of the pattern graph. We first define a tree decomposition and then its nice counterpart.
\begin{definition}
A \emph{tree decomposition} of a graph~$F$ is a triple $(T, \beta, r)$, where~$T$ is a tree rooted at node~$r$ and  $\beta \colon V(T) \mapsto 2^{V(F)}$ is a mapping satisfying:
    (i) $\bigcup_{x \in V(T)} \beta(x) = V(F)$;
    (ii) $\forall \{u, v\} \in E(F)$ $\exists x \in V(T)$, such that $u, v \in \beta(x)$;
    (iii) $\forall u \in V(F)$ the nodes $\{x \in V(T) \mid u \in \beta(x)\}$ form a connected subtree of~$T$.
\end{definition}

We shall denote bag~$\beta(x)$ as~$\mathcal{V}_x$. The width of tree decomposition $(T, \beta, r)$ is $\max_{x \in V(T)} |\mathcal{V}_x|-1$. Treewidth $\sc{tw}(F)$ of graph~$F$ is 
the minimal width of a~tree decomposition of~$F$ over all such decompositions.

\begin{definition}
A tree decomposition of a graph~$F$ is \emph{nice} if $\deg_T(r) = 1$, $\mathcal{V}_r = \emptyset$, and 
each node $x \in V(T)$ is of one of the following four types:
\begin{itemize}
  \item \emph{Leaf node}---$x$ has no children and $|\mathcal{V}_x| = 1$;
  \item \emph{Introduce node}---$x$ has exactly one child~$y$ and $\mathcal{V}_x = \mathcal{V}_y \cup \{u\}$ for some $u \in V(F) \setminus \mathcal{V}_y$;
  \item \emph{Forget node}---$x$ has exactly one child~$y$ and $\mathcal{V}_x = \mathcal{V}_y \setminus \{u\}$ for some $u \in \mathcal{V}_y$;
  \item \emph{Join node}---$x$ has exactly two children~$y, z$ and $\mathcal{V}_x = \mathcal{V}_y = \mathcal{V}_z$.
\end{itemize}
\end{definition}
Note that for practical purposes, we use a slightly modified definition of nice tree decomposition in this paper. As the algorithm starts the computation in a leaf node, using the standard definition with empty bags of leaves~\cite{kniha} would imply that the tables for leaves would be somewhat meaningless and redundant.
Therefore, we make bags of leaf nodes contain a single vertex.

\begin{definition}\label{def_v_star}
For a tree decomposition $(T, \beta, r)$, we denote by~$\mathcal{V}_x^*$ the set of vertices in~$\mathcal{V}_x$ and in~$\mathcal{V}_y$ for all descendants~$y$ of~$x$ in~$T$. Formally
$\mathcal{V}_x^* = \mathcal{V}_x \cup \bigcup_{y \text{ is a descendant of } x \text{ in } T} \mathcal{V}_y$.
\end{definition}
Note that, by Definition \ref{def_v_star}, for the root~$r$ of~$T$ we have $\mathcal{V}_r^* = V(F)$ and $F[\mathcal{V}_r^*] = F$.

\section{Algorithm Description}
\label{sec:algorithm}
In this section we first briefly describe the idea of the original color coding algorithm~\cite{ayz}, show, how to alter the computation in order to reduce its time and memory requirements, and describe implementation details and further optimizations of the algorithm.

\subsection{Idea of the Algorithm}
\label{subs:idea}

The critical idea of color coding is to reduce the problem to its colorful version. For a graph~$G$ and a pattern graph~$F$, we color the vertices of~$G$ with exactly~$n_{F}$ colors. We use the randomized version, i.e., we create a random
coloring $\zeta \colon V(G) \mapsto \{1, 2, \ldots, n_{F}\}$. After the coloring, the algorithm considers as valid only subgraphs~$G'$ of~$G$ that are colorful copies of~$F$ as follows. 

\begin{definition}
Subgraph~$G'$ of a graph~$G$ is a \emph{colorful copy} of~$F$ with respect to coloring~$\zeta \colon V(G) \mapsto \{1, 2, \ldots, n_F\}$, if~$G'$ is isomorphic to~$F$ and all of its vertices 
are colored by distinct colors in~$\zeta$.
\end{definition}

As the output of the algorithm heavily depends on the chosen random coloring of~$G$, in order to reach some predefined success rate of the algorithm, we need to repeat the process of coloring several times.
The probability of a particular occurrence of pattern graph~$F$ becoming colorful with respect to the random coloring is $\frac{n_{F}!}{n_{F}^{n_{F}}}$, which tends to $e^{-n_{F}}$ for large~$n_{F}$.
Therefore, by running the algorithm $\mathrm{e}^{n_{F}\log{\frac{1}{\varepsilon}}}$ times, each time with a random coloring~$\zeta \colon V(G) \mapsto \{1, 2, \ldots, n_{F}\}$, the probability that an existing occurrence of the pattern will be revealed in none of the runs
is at most~$\varepsilon$. 
While using more colors can reduce the number of iterations needed, it also significantly increases the memory requirements. Hence, we stick to $n_F$ colors.
Even though it is possible to derandomize such algorithms, e.g., by the approach shown in~\cite{kniha}, 
in practice the randomized approach usually yields the results much quicker, as discussed in~\cite{pcc}.
Moreover, we are not aware of any actual implementation of the derandomization methods.

The main computational part of the algorithm is a dynamic programming.
The target is to create a graph isomorphism $\Phi \colon V(F) \mapsto V(G)$. We do so by traversing the
nice tree decomposition $(T, \beta, r)$ of the pattern graph~$F$ and at each node $x \in V(T)$ of the tree decomposition, we construct possible partial mappings $\varphi \colon \mathcal{V}^*_x \to V(G)$
with regard to required colorfulness of the copy.
Combination of partial mappings consistent in colorings then forms a desired resulting mapping.

The semantics of the dynamic programming table is as follows. For any tree decomposition node $x \in V(T)$, any partial mapping $\varphi \colon \mathcal{V}_x \mapsto V(G)$ and any color subset 
$C \subseteq \{1, 2, \ldots, n_{F}\}$, we define $\mathcal{D}(x, \varphi, C) = 1$ if there is an isomorphism~$\Phi$ of $F[\mathcal{V}_x^*]$ to a subgraph~$G'$ of~$G$ such that:
\begin{enumerate}[(i)]
  \item for all $u \in \mathcal{V}_x, \Phi(u) = \varphi(u)$;
  \item $G'$ is a colorful copy of $F[\mathcal{V}_x^*]$ using exactly the colors in~$C$, that is, $\zeta(\Phi(\mathcal{V}_x^*)) = C$ and $\zeta$ is injective on $\Phi(\mathcal{V}_x^*)$.
\end{enumerate}
If there is no such isomorphism, then we let $\mathcal{D}(x, \varphi, C) = 0$. 
We denote all configurations $(x, \varphi, C)$ for which $\mathcal{D}(x, \varphi, C) = 1$ as \emph{nonzero} configurations.

The original version of the algorithm is based on top-down dynamic programming approach with memoization of already computed results.
That immediately implies a big disadvantage of this approach---it requires the underlying dynamic programming table (which is used for memoization) 
to be fully available throughout the whole run of the algorithm. 
To avoid this inefficiency in our modification we aim to store only nonzero configurations, similarly to the recent ``positive instance driven'' dynamic programming approach~\cite{Tamaki17}.

\section{Obtaining a Nice Tree Decomposition}
\label{sec:obtaining}

The algorithm requires a nice tree decomposition of the pattern graph for its work. 
The running time of the algorithm actually does not depend on the treewidth of the graph, but rather on the width of the tree decomposition supplied.
On one hand, for
a given graph~$G$ and an integer~$k$, the problem of determining whether the treewidth of~$G$ is at most~$k$, is \npc~\cite{arn}.
On the other hand, as the main algorithm is exponential in the size of the pattern anyway, we can afford to use exponential time algorithms in order to obtain a tree decomposition of minimum possible width.
We employ
known technique based on graph triangulation, chordal graphs and elimination orderings, combining the ideas of Bodlaender et al.~\cite{tw_exact,tw_upper}.
Time complexity of this algorithm is $\mathcal{O}(n^2_F \cdot 2^{n_F})$.

After obtaining tree decomposition of the pattern graph, we construct a nice tree decomposition as described by Cygan et al.~\cite{kniha}---any tree decomposition of a graph~$F$ that consists 
of $\mathcal{O}(n_F)$ nodes with width at most~$t$, can be, in $\mathcal{O}(t^2 \cdot n_{F})$ time, transformed to a nice tree decomposition of~$F$ with $\mathcal{O}(t \cdot n_F)$ nodes and width bounded by~$t$.
Our further modification to the nice tree decomposition does not affect the running time.

\subsection{Initial Algorithm Modification}

In our implementation, we aim to store only nonzero configurations, therefore we need to be able to construct nonzero configurations
of a~parent node just from the list of nonzero configurations in its child/children.

We divide the dynamic programming table~$\mathcal{D}$ into lists of nonzero configurations, where each nice tree decomposition node has a list of its own. 
Formally, for every node $x \in V(T)$, let us denote by~$\mathcal{D}_{x}$ a list of all mappings~$\varphi$ with a list of their corresponding color sets~$C$, for which $\mathcal{D}(x, \varphi, C) = 1$. 
The list~$\mathcal{D}_{x}$ for all $x \in V(T)$ is, in terms of contained information, equivalent to maintaining the whole table~$\mathcal{D}$---all configurations not
present in the lists can be considered as configurations with a result equal to zero.

\subsubsection*{Dynamic Programming Description}

We now describe how to compute the lists $\mathcal{D}(x, \varphi, C)$ for each type of a nice tree decomposition node.

For a \emph{leaf} node $x \in T$, there is only a single vertex~$u$ in~$\mathcal{V}_x^*$ to consider. We can thus map~$u$ to all possible vertices of~$G$, and we obtain a list with~$n_{G}$ partial mappings~$\varphi$, 
in which the color list for each mapping contains a~single color set~$\{\zeta(\varphi(u))\}$. 

For an \emph{introduce} node $x \in T$ and its child~$y$ in~$T$, we denote by~$u$ the vertex being introduced in~$x$, i.e., $\{u\} = \mathcal{V}_x \setminus \mathcal{V}_y$. For all nonzero combinations of a~partial mapping
and a color set $(\varphi ', C')$ in the list~$\mathcal{D}_{y}$, we try to extend~$\varphi '$ by all possible mappings of the vertex~$u$ to the vertices of~$G$. We denote one such
a mapping as~$\varphi$. We can consider mapping~$\varphi$ as correct, if 
(i) the new mapping~$\varphi(u)$ of the vertex~$u$ extends the previous colorset~$C'$, that is, $C = C' \cup \{\zeta(\varphi(u))\} \neq C'$, and (ii) $\varphi$ is \emph{edge consistent}, that is, for all 
edges $\{v, w\} \in E(F)$ between currently mapped vertices, i.e., in our case $v, w \in \mathcal{V}_x$, there must be an edge $\{\varphi (v), \varphi (w)\} \in E(G)$. However, because~$\varphi '$ was by construction 
already edge consistent, it suffices to check the edge consistency only for all edges in $F[\mathcal{V}_x]$ with~$u$ as one of their endpoints, i.e., for all edges $\{u, w\} \in E(F[\mathcal{V}_x])$ with 
$w \in N_{F[\mathcal{V}_x]}(u)$. After checking those two conditions, we can add $(\varphi, C)$ to $\mathcal{D}_{x}$.

For a \emph{forget} node $x \in V(T)$ and its child~$y$ in~$T$, there is not much work to do, as in the bottom-up construction, we directly obtain the parent list of nonzero configurations. 
We denote by~$u$, the vertex being forgotten in~$x$, i.e., $\{u\} = \mathcal{V}_y \setminus \mathcal{V}_x$. 
In this case, for all partial mappings~$\varphi '$ in the list~$\mathcal{D}_{y}$, 
we create a~new mapping~$\varphi$ that excludes the mapping~$\varphi'(u)$ for vertex~$u$, i.e., $\varphi = \restr{\varphi '}{\mathcal{V}_x}$. Color sets corresponding to~$\varphi '$
are carried over to~$\varphi$, as they represent colors already used in the construction. After this step, we might need to merge color lists of previously different mappings, as after the removal of the mapping~$\varphi'(u)$, they
might become the same mappings. 

For a \emph{join} node $x \in V(T)$, we denote by~$y$ and~$z$ its children in~$T$. We traverse the children lists~$\mathcal{D}_{y}$ and~$\mathcal{D}_{z}$ and look for partial mappings~$\varphi'$ and~$\varphi''$, 
for which $\varphi' = \varphi''$ holds. Such mappings are the only ones to potentially form a new nonzero configuration in the parent list, as due to the fact that $\mathcal{V}_x = \mathcal{V}_y = \mathcal{V}_z$,
we construct the new partial mapping~$\varphi$ as $\varphi = \varphi ' = \varphi ''$. However, for each
such mapping~$\varphi$, we must also construct the new list of color sets, which would afterwards be corresponding to the mapping in the parent list. We do that by traversing color lists
corresponding to mappings~$\varphi'$ and~$\varphi''$ in~$\mathcal{D}_{y}$ and~$\mathcal{D}_{z}$, respectively, and for particular sets~$C'$ and~$C''$ from the color lists of~$\varphi '$ and~$\varphi ''$
construct a new color set $C = C' \cup C''$. We must check, whether the intersection of~$C'$ and~$C''$ contains exactly the colors to color the vertices in~$\mathcal{V}_x$. That is, for mapping~$\varphi$, we add to~$\mathcal{D}_{x}$ a color set $C= C' \cup C''$, if $C' \cap C'' = \{\zeta(\varphi(\mathcal{V}_x))\}$.

Because we build the result from the leaves
of the nice tree decomposition, we employ a recursive procedure on its root, in which we perform the computations in a way of a post-order traversal of a tree. From each visited node, we obtain a bottom-up dynamic 
programming list of nonzero configurations. After the whole nice tree decomposition is traversed, we obtain a list of configurations, that were valid in its root. Such configurations thus represent solutions 
found during the algorithm, from which we afterwards reconstruct results. Note that as we prepend a root with no vertices in its bag to the nice tree decomposition, there is a nonzero number of solutions
if and only if, at the end of the algorithm, the list~$\mathcal{D}_{r}$ contains a single empty mapping using all colors.

\subsection{Further Implementation Optimizations}

\subsubsection*{Representation of Mappings}
For mapping representation, we suppose that the content of all bags of the nice tree decomposition stays in the same order during the whole algorithm. 
This natural and easily satisfied condition allows us to represent a mapping 
$\varphi \colon \mathcal{V}_x \mapsto V(G)$ in a nice tree decomposition node~$x$ simply by an ordered tuple of~$|\mathcal{V}_x|$ vertices from~$G$.
From this, we can easily determine which vertex from~$F$ is mapped to 
which vertex in~$G$. Also, for a mapping in an~introduce or a~forget node, we can describe a position in the mapping, on which the process of introducing/forgetting takes place.

\subsubsection*{Representation of Color Sets}
We represent color sets as bitmasks, where the $i$-th bit states whether color~$i$ is contained in the set or not. For optimization purposes, we represent bitmasks with an integer number. As we use~$n_{F}$ colors in the algorithm 
 and restricted ourselves to pattern graphs with at most $32$ vertices, we represent a~color set with a~$32$-bit number.

\subsubsection*{Compressing the Lists}
Because we process the dynamic programming lists one mapping at a time, we store these lists in a compressed way and decompress them only on a mapping retrieval basis.
We serialize dynamic programming lists into a simple buffer of bytes. We store a dynamic programming list as a continuous group of records, each of which represents one partial mapping and its corresponding list of color sets. 
Each record contains:
\begin{itemize}
  \item a mapping in the form of ordered tuple of vertices,
  \item the number of color sets included,
  \item and color sets corresponding to the mapping in the form of non-negative integer numbers.
\end{itemize}
While deserializing, the number of vertices in a~mapping can be easily determined by the size 
of the bag of a particular node.

As there is no requirement on the order of the color sets,
we store these sets sorted in the increasing order of their number representation. This allows us to effectively use delta compression.
Moreover, we use variable length encoding to store the numbers into buffer.

For several routines we use the LibUCW library,\footnote{LibUCW is downloadable from \url{http://www.ucw.cz/libucw/}.}
a C language library highly optimized for performance.
The employed routines include data structures like growing buffers, hash tables, red-black trees,
fast sorters, fast buffered input/output, and also the efficient variable length encoding of integers.

\subsubsection*{Masking Unprofitable Mappings}

Our implementation supports an extended format of input graphs where one can specify for each vertex of the network, which vertices of the pattern can be mapped to it.
This immediately yields a simple degree-based optimization. Before the run of the main algorithm, we perform a linear time preprocessing of input graphs
and only allow a vertex $y \in V(F)$ to be mapped to a vertex $x \in V(G)$ if $\deg_G(x) \geq \deg_F(y)$.

\subsubsection*{Mapping Expansion Optimizations}

The main ``brute-force'' work of the algorithm is performed in two types of nodes---leaf and introduce nodes, as we need to try all possible mappings of
a~particular vertex in a~leaf node or all possible mappings of an~introduced vertex in a~introduce node to a vertex from~$G$.
We describe ways to optimize the work in introduce nodes in this paragraph.

Let~$x$ be an introduce node, $u$ the vertex introduced and~$\varphi$ a mapping  from a nonzero configuration for the child of~$x$.
We always need to check whether the new mapping of~$u$ is edge consistent with the mapping~$\varphi$ of the remaining vertices for the corresponding bag, i.e., whether all edges of~$F$ incident on~$u$ would be realized by an edge in~$G$.
Therefore, if~$u$ has any neighbors in $F[\mathcal{V}_x]$, then a vertex of~$G$ is a candidate for the mapping of~$u$ only if it is a neighbor of all vertices in the set $\varphi(N_{F[\mathcal{V}_x]}(u))$, i.e., the vertices of~$G$, where the neighbors of~$u$ in~$F$ are mapped.
Hence, we limit the number of candidates by using the adjacency lists of the already mapped vertices.

In the case $\deg_{F[\mathcal{V}_x]}(u) = 0$ we have to use different approach.
The pattern graphs~$F$ tend to be smaller than the input graphs~$G$ by several orders of magnitude.
Hence, if the introduced vertex is in the same connected component of~$F$ as some vertex already present in the bag, a~partial mapping processed in an introduce node 
anchors the possible resulting component to a certain position in~$G$. 
During the construction of possible mapping of~$u$, it is useless to try mapping~$u$ to vertices in~$G$ that are very distant to the current position and, therefore, could
by no means form a~resulting subgraph isomorphic to the component of~$F$. 
We obtain the maximal possible distance to be considered in~$G$ as a minimal distance of~$u$ to a vertex in $\mathcal{V}_x \setminus \{u\}$.
I.e., we determine $w= \mathrm{argmin}_{v \in \mathcal{V}_x \setminus \{u\}}\mathrm{dist}_F(v,u)$.
Due to the limit on pattern graph size, we precompute distances between every pair of vertices by using BFS on each vertex before the start of the algorithm.
Then it suffices to try vertices from~$G$ that are in distance at most $\mathrm{dist}(w,u)$ from~$\varphi(w)$ in~$G$; again, by a simple BFS usage.

Only if there is no vertex in the bag sharing a connected component of~$F$ with~$u$, we have to fall back to trying all possible mappings.

\section{Result Reconstruction}
\label{sec:reconstruction}

It is usual in dynamic programming to store a witness or all witnesses for each reasonable value in the table. However, this would completely neglect the effect of compression of the lists and significantly increase the memory requirements.

Hence, to enumerate all results, we recursively traverse the computed dynamic programming lists~$\mathcal{D}_{x}$ starting with the root~$r$ of underlying nice tree decomposition.
The main idea is to ask the child~$y$ (or children $y, z$) of a node, how was a certain partial mapping~$\varphi$ obtained during the computation.
In other words, for each partial mapping~$\varphi$ in~$\mathcal{D}_{x}$ of interest, we will extract partial mappings~$\varphi'$ in~$\mathcal{D}_{y}$ (or possibly also partial mappings~$\varphi''$ in~$\mathcal{D}_{z}$) that lead to the addition of~$\varphi$ to~$\mathcal{D}_{x}$.
To preserve the colorfulness (and thus validity) of glued partial mappings, with a partial mapping~$\varphi$ we also recursively pass a set of colors that can still be used in the choice of corresponding partial mappings to glue with~$\varphi$.
By applying this approach recursively and by gluing the partial information together in mappings, we obtain all possible ways an empty mapping in~$r$ could have been obtained---which is exactly the same as all possible ways a pattern graph can be mapped to the original graph.
For the efficiency of the computation, we process each node only once and we thus recursively pass all partial mappings of interest and their remaining colorsets at once.

Formally, we will consider a recursive call of function $\mathcal{R}(x, \mathcal{M})$, where $x \in V(T)$, $\mathcal{M}$ is a set of pairs $(\varphi, C)$, $\varphi$ is a partial mapping $\varphi \colon \mathcal{V}_x \mapsto V(G)$ and~$C$ is a color subset 
$C \subseteq \{1, 2, \ldots, n_{F}\}$. This function returns a list~$\mathcal{L}$ containing, for each pair $(\varphi, C)$ in~$\mathcal{M}$, a list $L_{(\varphi, C)}$ of mappings $\Phi : \mathcal{V}^*_x \mapsto V(G)$ that lead to the appearance of $(\varphi, C)$ in~$\mathcal{D}_{x}$. 

In detail, the function $\mathcal{R}(x, \mathcal{M})$ proceeds as follows. If node~$x$ has a child~$y$, we first scan~$D_{t,y}$ for pairs $(\varphi', C')$,
in which~$\varphi'$ and~$\varphi$ and colorsets~$C'$ and~$C$ are in some sense consistent.
Pairs satisfying these conditions are then recursively passed to~$y$ in a list~$\widehat{\mathcal{M}}$, from which we obtain a list~$\widehat{\mathcal{L}}$.
In case of a join node, we perform this process for both children.
After that, we construct the resulting list from recursively computed list(s) by an addition of a mapping of the introduced vertex (in the case of a introduce node) or by all possible combinations of mappings (in the case of a join node). 
In the end, we sometimes need to flatten the resulting list, so all resulting partial mappings corresponding to a pair $(\varphi, C)$ are in a single list $L_{(\varphi, C)} \in \mathcal{L}$.
We now describe the process and consistency conditions formally for each of the node types.

For a \emph{leaf} node $x \in T$, we simply return a list~$\mathcal{L}$ of single element lists~$L_{(\varphi, C)}$ for each pair $(\varphi, C) \in \mathcal{M}$. Each such list contains a partial mapping $\Phi \assign \varphi$.

For an \emph{introduce} node $x \in T$ and its child~$y$ in~$T$, we denote by~$u$ the vertex being introduced in~$x$, i.e., $\{u\} = \mathcal{V}_x \setminus \mathcal{V}_y$. 
We create~$\widehat{\mathcal{M}}$ consisting of $(\widehat{\varphi}, \widehat{C})$ by setting, for each pair $(\varphi, C) \in \mathcal{M}$,
$\widehat{\varphi} \assign \restr{\varphi}{\mathcal{V}_y}$ and $\widehat{C} \assign C \setminus \{\zeta(\varphi(u))\}$. 
Note that $(\widehat{\varphi}, \widehat{C})$ must be in~$\mathcal{D}_{y}$, as otherwise $(\varphi, C)$ would not be present in~$\mathcal{M}$ (or equivalently in~$\mathcal{D}_{x}$).
Then we obtain $\widehat{\mathcal{L}} \assign \mathcal{R}(y, \widehat{\mathcal{M}})$.
Let $(\widehat{\varphi}, \widehat{C}) \in \widehat{\mathcal{M}}$ be constructed from $(\varphi, C) \in \mathcal{M}$ and~$\widehat{L}_{(\widehat{\varphi}, \widehat{C})}$ be the corresponding list of~$\widehat{\mathcal{L}}$.
We construct the list~$L_{(\varphi, C)}$ corresponding to $(\varphi, C)$ by extending each $\Phi \in \widehat{L}_{(\widehat{\varphi}, \widehat{C})}$ by~$\varphi(u)$.

For a \emph{forget} node $x \in T$ and its child~$y$ in~$T$, we denote by~$u$ the vertex being forgotten in~$x$, i.e., $\{u\} = \mathcal{V}_y \setminus \mathcal{V}_x$. 
We add to~$\widehat{\mathcal{M}}$ for each $(\varphi, C) \in \mathcal{M}$ all $(\widehat{\varphi}, \widehat{C})$ from~$\mathcal{D}_{y}$ such that $\varphi = \restr{\widehat{\varphi}}{\mathcal{V}_x}$ and $\widehat{C} = C$. 
Then we obtain $\widehat{\mathcal{L}} \assign \mathcal{R}(y, \widehat{\mathcal{M}})$.
We construct~$\mathcal{L}$ from~$\widehat{\mathcal{L}}$ as follows. We do not need to modify any~$\Phi$ in any $\widehat{L} \in \widehat{\mathcal{L}}$, as the computation in forget node adds no new information to the mapping.
However, as there may have been multiple (say~$k$) pairs $(\widehat{\varphi_1}, \widehat{C_1}), \ldots, (\widehat{\varphi_k}, \widehat{C_k}) \in \widehat{\mathcal{M}}$ constructed from a single $(\varphi, C)$, we flatten all lists $L_{(\widehat{\varphi_1}, \widehat{C_1})}, \ldots, L_{(\widehat{\varphi_k}, \widehat{C_k})} \in \widehat{\mathcal{L}}$ obtained from such pairs into a single list~$L_{(\varphi, C)}$ corresponding to $(\varphi, C) \in \mathcal{M}$.

For a \emph{join} node $x \in V(T)$, we denote by~$y$ and~$z$ its children in~$T$. 
In this case, we create~$\mathcal{M}'$ and~$\mathcal{M}''$ consisting of $(\varphi', C')$ and $(\varphi'', C'')$, respectively.
For each $(\varphi, C)$, we find all $(\varphi', C') \in \mathcal{D}_{y}$ and $(\varphi'', C'') \in \mathcal{D}_{z}$, such that $\varphi = \varphi' = \varphi''$, $C = C' \cup C''$ and $C' \cap C'' = \{\zeta(\varphi(\mathcal{V}_x))\}$ and add them to~$\mathcal{M}'$ and~$\mathcal{M}''$, respectively.
Then we obtain $\mathcal{L}' \assign \mathcal{R}(y, \mathcal{M}')$ and $\mathcal{L}'' \assign \mathcal{R}(z, \mathcal{M}'')$.
Let $(\varphi_1', C'_1)$ and $(\varphi_1'', C''_1)$, \dots, $(\varphi_k', C'_k)$ and $(\varphi_k'', C''_k)$ be the pairs constructed from single $(\varphi, C) \in \mathcal{M}$
and let $L_{(\varphi_1', C'_1)}, \ldots, L_{(\varphi_k', C'_k)}$ and $L_{(\varphi_1'', C''_1)}, \ldots, L_{(\varphi_k'', C''_k)}$ be the corresponding lists in~$\mathcal{L}'$ and~$\mathcal{L}''$, respectively.
We obtain~$L_{(\varphi, C)}$ corresponding to~$(\varphi, C)$ as a union of lists $L_1, \ldots, L_k$, where~$L_i$ contains the union of~$\Phi'$ and~$\Phi''$ for each $\Phi' \in L_{(\varphi_i', C'_i)}$ and each $\Phi''  \in L_{(\varphi_i'', C''_i)}$.

\section{Experimental Results}
\label{sec:experiments}

The testing was performed on a 64-bit linux system with Intel Xeon CPU E3-1245v6@3.70GHz and 32GB 1333MHz DDR3 SDRAM memory. 
The module was compiled with \verb|gcc| compiler (version 7.3.1)
with \verb|-O3| optimizations enabled. 
Implementation and instances utilized in the testing are available at \url{http://users.fit.cvut.cz/malikjo1/subiso/}.
All results are an average of 5 independent measurements.

We evaluated our implementation in several ways.
Firstly, we compare available implementations on two different real world source graphs and a set of more-or-less standard target graph patterns.
Secondly, we compare available implementations on instances from ICPR2014 Contest on Graph Matching Algorithms for Pattern Search in Biological Databases \cite{ICPR2014} with suitably small patterns.
We also adapt the idea of testing the algorithms on Erd\H{o}s-R{\' e}nyi random graphs~\cite{McCreeshPST18}.

\subsection{Algorithm Properties and Performance}
In the first two subsection we used two different graphs of various properties as target graph~$G$. 
The first instance, \sc{Images}, is built from an segmented image, and is a courtesy of~\cite{img}. 
It consists of $4838$ vertices and $7067$ edges. The second instance, \sc{Trans}, is a graph of transfers on bank accounts. It is a very sparse network,
which consists of $45733$ vertices and $44727$ undirected edges.

\begin{figure}[t]
  
  \centering
  \null\hfill
  \subfloat[Pattern $A$]
  {
      \vspace{-0.6cm}
      \begin{tikzpicture}
        [
        every node/.style={c}
        ]
        \node(0) at (0, 0) {};
        \node(1) at (0.5,0.5) {};
        \node(2) at (1,0) {};
        \node(3) at (0.75, -0.5) {};
        \node(4) at (1,0.5) {};
        \node(5) at (1.5,0.5) {};
        \node(6) at (1.5,0) {};
        \node(7) at (1.5,-0.5) {};
        \node(8) at (1.25,-1) {};
        
        \draw[-] (0) to (1);
        \draw[-] (1) to (2);
        \draw[-] (2) to (3);
        \draw[-] (2) to (4);
        \draw[-] (5) to (4);
        \draw[-] (6) to (4);
        \draw[-] (6) to (7);
        \draw[-] (7) to (8);
      \end{tikzpicture}
  }
  \hfill
  \subfloat[Pattern $B$]
  {
    \begin{tikzpicture}
      [
      every node/.style={c}
      ]
        \node(0) at (0, 0) {};
        \node(1) at (1,0) {};
        \node(2) at (0.5,0.72) {};
        \node(3) at (0, 0.72) {};
        \node(4) at (0.5,1.44) {};
        \node(5) at (1.5,0.5) {};
        \node(6) at (1.5,0) {};
        
        \draw[-] (0) to (1);
        \draw[-] (1) to (2);
        \draw[-] (0) to (2);
        \draw[-] (1) to (6);
        \draw[-] (1) to (5);
        \draw[-] (2) to (3);
        \draw[-] (3) to (4);
    \end{tikzpicture}
  }
  \hfill
  \subfloat[Pattern $C$]
  { 
    \begin{tikzpicture}
      [
      every node/.style={c}
      ]
      \node(0) at (-1, 0) {};
      \node(1) at (1,0) {};
      \node(2) at (0,1.44) {};
      
      \node(3) at (0, 0) {};
      \node(4) at (-0.5,0.72) {};
      \node(5) at (0.5,0.72) {};
      
      \node(6) at (1,1.44) {};
      \node(7) at (-1,1.44) {};
      \node(8) at (0,-0.72) {};
      
      \draw[-] (0) to (3);
      \draw[-] (0) to (4);
      \draw[-] (1) to (3);
      \draw[-] (1) to (5);
      \draw[-] (2) to (4);
      \draw[-] (2) to (5);
      \draw[-] (6) to (1);
      \draw[-] (6) to (2);
      \draw[-] (7) to (0);
      \draw[-] (7) to (2);
      \draw[-] (8) to (0);
      \draw[-] (8) to (1);
    \end{tikzpicture}
  }
  \hfill
  \subfloat[Pattern $D$]
  { 
    \begin{tikzpicture}
      [
      every node/.style={c}
      ]
      \def \radius {0.7cm}
      \def\n{5}
      
      \node(p1) at (-1.5,0.5) {};
      \node(p2) at (-1.5,-0.5) {};
      \draw[-] (p1) to (p2);
      
      \foreach \i in {1,...,\n}
      {
        \node(\i) at ({360/\n * (\i + 2) + 90/\n}:\radius) {};
      }
      \pgfmathtruncatemacro{\last}{\n-1}
      \foreach \i in {1,...,\last}
      {
        \pgfmathtruncatemacro{\nxt}{\i+1}
        \draw[-] (\i) to (\nxt);
      }
      \draw[-] (1) to (5);
    \end{tikzpicture}
  }
  \hfill\null
  \caption{Pattern graph illustration}\label{figpatt}
\end{figure}
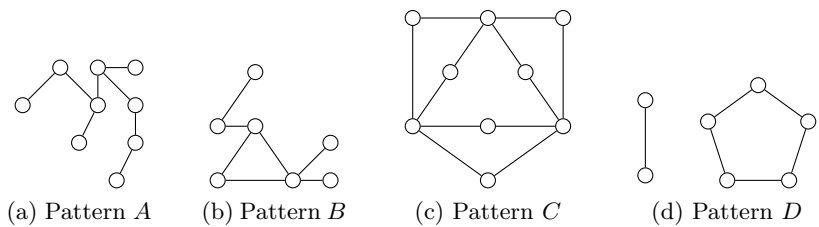

For the pattern graphs, we first use a standard set of basic graph patterns, as the treewidth of such graphs is well known and
allows a clear interpretation of the results. In particular, we use paths, stars, cycles, an complete graphs on $n$  vertices, denoted $P_n$, $S_n$, $C_n$, and $K_n$ with treewidth 1, 1, 2, and $n-1$, respectively. We further used grids $G_{n,m}$ on $n \times m$ vertices, with treewidth $\mathrm{min}\{n, m\}$.
Secondly, we use a special set of pattern graphs in order to demonstrate performance on various patterns.
Patterns $A$, $B$, $C$, and $D$ have $9$, $7$, $9$, and $7$ vertices, $8$, $7$, $12$, $6$ edges, and treewidth $1$, $2$, $2$, and $2$, respectively. Patterns $A$, $B$, and $D$ appear in both dataset, pattern $C$ in neither and pattern $D$ is disconnected. 
Description of these pattern graphs is shown in \autoref{prop} and their illustration is shown in \autoref{figpatt}.
\begin{table}[t]
 \centering
 \caption[Pattern graphs]{Pattern graphs.}
 \begin{tabular}{c||c|c}
 \label{prop}
Pattern $F$ & $\sc{tw}(F)$ & Description\\
\hline \hline
$P_n$ & $1$ & path on $n$ vertices \\ \hline
$S_n$ & $1$ & star on $n$ vertices\\ \hline
$C_n$ & $2$ & cycle on $n$ vertices\\ \hline
$K_n$ & $n - 1$ & complete graph on $n$ vertices\\ \hline
$G_{n,m}$ & $\mathrm{min}\{n, m\}$ & grid graph on $n \times m$ vertices\\ \hline
Pattern $A$ & $1$ & subgraph of both datasets with $9$ vertices, $8$ edges \\ \hline
Pattern $B$ & $2$ & subgraph of both datasets with $7$ vertices, $7$ edges \\ \hline
Pattern $C$ & $2$ & graph with $9$ vertices, $12$ edges \\ \hline
Pattern $D$ & $2$ & disconnected graph with $7$ vertices, $6$ edges \\ \hline          
 \end{tabular}                                                                  
\end{table}

Due to randomization, in order to achieve some preselected constant error rate, we need to repeat the computation more than once. The number of found results thus depends not only on the quality of the
algorithm, but also on the choice of the number of its repetitions. Hence, it is logical to measure performance of the single run of the algorithm. Results from such a testing, however, 
should be still taken as a rough average, because the running time of a single run of the algorithm depends on many factors. 

Therefore, we first present measurements, where we average the results of many single runs of the algorithm. We average not only the time and space needed, but also the number of found subgraphs.
To obtain the expected time needed to run the whole algorithm, it suffices to sum the time needed to create a nice tree decomposition and~$\ell$ times the time required for a single run, if there are~$\ell$ runs in total.

\begin{table}[t]
 \centering
 \caption[Performance of a single run of the algorithm on \sc{Images} dataset]{Performance of a single run of the algorithm on \sc{Images} dataset.}
 \vspace{10px}
 \begin{tabular}{c||c|c||c|c||c}
 \label{res_img}
Pattern	& Comp. time [ms] & Comp. memory [MB] & Occurrences\\
\hline \hline
$\mathcal{P}_5$    & 240   & 12.73 & 3488.21 \\ \hline
$\mathcal{P}_{10}$ & 160 & 8.52 & 732.46 \\ \hline
$\mathcal{P}_{15}$ & 90 & 10.54 & 76.18 \\ \hline
$\mathcal{S}_5$    & 4   & 5.37 & 114.72 \\ \hline
$\mathcal{C}_5$    & 20   & 7.24 & 239.17 \\ \hline
$\mathcal{C}_{10}$ & 70   & 9.34 &  26.64 \\ \hline
$\mathcal{K}_{4}$  & 5   & 6.46 & 0 \\ \hline
$G_{3,3}$  & 90   & 13.42 &  0 \\ \hline
Pattern $A$  & 80   & 9.14 &  292.48 \\ \hline
Pattern $B$  & 10   & 7.17 &  6.85 \\ \hline
Pattern $C$  & 10   & 5.30 &  0 \\ \hline
Pattern $D$  & 40   & 10.14 &  426.76 \\ \hline
 \end{tabular}
\end{table}

\begin{table}[t]
 \centering
 \caption[Performance of a single run of the algorithm on \sc{Trans} dataset]{Performance of a single run of the algorithm on \sc{Trans} dataset.}
 \vspace{10px}
 \begin{tabular}{c||c|c||c|c||c}
 \label{res_Trans}
Pattern	& Comp. time [s] & Comp. memory [MB] & Occurrences\\
\hline \hline
$\mathcal{P}_5$   & 6.15 & 33.82 & 54572.94 \\ \hline
$\mathcal{P}_{10} $ & 0.32 & 34.40 & 562.54 \\ \hline
$\mathcal{P}_{15} $ & 0.17 & 34.83 & 91.49 \\ \hline
$\mathcal{C}_5$ & 0.11 & 30.11 & 15.32 \\ \hline
$\mathcal{C}_{10}$ & 0.24 & 33.84 &  0 \\ \hline
$\mathcal{K}_{4}$ & 0.09 & 27.91 &  0.72 \\ \hline
$G_{3,3}$  & 2.16   & 48.68 &  0 \\ \hline
Pattern $A$  & 54.42   & 109.88 &  94145.82 \\ \hline
Pattern $B$  & 0.18   & 34.12 &  1127.11 \\ \hline
Pattern $C$  & 0.29   & 35.16 &  0 \\ \hline
Pattern $D$  & 0.25   & 47.54 &  178.29 \\ \hline
 \end{tabular}
\end{table}

\subsection{Comparison on Real World Graphs and Fixed Graph Patterns}
\label{sec:real_word}

We compare our implementation to three other tools for subgraph enumeration: RI algorithm~\cite{ri} (as implemented in \cite{leskovec2016snap}), LAD algorithm~\cite{lad} and color coding algorithm for weighted path enumeration~\cite{sig} 
(by setting, for comparison purposes, all weights of edges to be equal). The comparison is done on the instances from previous subsection and only on pattern graphs which occur at least once in a particular target graph.

In comparison, note the following specifics of measured algorithms. 
The RI algorithm does not support outputting solutions, which might positively affect its performance.
LAD algorithm uses adjacency matrix to store input graphs, and thus yields potentially limited use for graphs of larger scale.
Neither of RI or LAD algorithms supports enumeration of disconnected patterns.\footnote{
When dealing with disconnected patterns, one could find the components of the pattern one by one, omitting the vertices of the host graph used by the previous component.
However, this would basically raise the running time of the algorithm to the power equal to the number of components of the pattern graph.}
Also we did not measure the running time of the weighted path algorithm on non-path queries and also on \sc{Trans} dataset, as its implementation
is limited to graph sizes of at most $32\,000$.

We run our algorithm repeatedly to achieve an error rate of $\varepsilon = \frac{1}{e}$.
In order to be able to measure the computation for larger networks
with many occurrences of the pattern, we measure only the time required to retrieve no more than first $100\,000$ solutions
and we also consider running time greater than 10 minutes (600 seconds) as a timeout.
Since we study non-induced occurrences (and due to automorphisms) there might be several ways to map the pattern to the same set of vertices.
Other measured algorithms do count all of them.
Our algorithm can behave also like this, or can be switched to count only occurrences that differ in vertex sets.
For the sake of equal measurement, we use the former version of our algorithm.

\begin{table}[t]
 \centering
 \caption[Comparison of running time on \sc{Images} dataset (in seconds)]{Comparison of running time on \sc{Images} dataset (in seconds).}
 \vspace{10px}
 \begin{tabular}{c||c|c|c|c}
 \label{c_i}
Pattern	& Our algorithm & RI algorithm & LAD algorithm & Weighted path\\
\hline \hline
$\mathcal{P}_5$    & 31.12   & 0.11 & 28.86 & 362.41\\ \hline
$\mathcal{P}_{10}  $ & 53.17 & 1.25 & 13.63 & $>600$\\ \hline
$\mathcal{P}_{15}  $ & 104.30 & 3.7  & 8.18 & $>600$\\ \hline
$\mathcal{S}_5$    & 0.94   & 0.07 & 0.43 & --\\ \hline
$\mathcal{C}_5$    & 4.98   & 0.14 & 35.18 &--\\ \hline
$\mathcal{C}_{10}$ & 151.25   & 3.44 & 174.27 &--\\ \hline
Pattern $A$  & 43.11 & 0.82 & 36.60 &--\\ \hline
Pattern $B$  & 91.93   & 0.41 &  0.83 &--\\ \hline
Pattern $D$  & 23.54   & -- &  -- &--\\ \hline 
 \end{tabular}
\end{table}

\begin{table}[t]
 \centering
 \caption[Comparison of running time on \sc{Trans} dataset (in seconds)]{Comparison of running time on \sc{Trans} dataset (in seconds).}
 \vspace{10px}
 \begin{tabular}{c||c|c|c|c}
 \label{c_t}
Pattern	& Our algorithm & RI algorithm & LAD algorithm & Weighted path\\
\hline \hline
$\mathcal{P}_5$    & 11.64  & 2.53 & 59.57 & --\\ \hline
$\mathcal{P}_{10}  $ & 44.72 & 4.77  & 34.00 & --\\ \hline
$\mathcal{P}_{15}  $ & 295.11 & 24.11 & 28.98 & --\\ \hline

$\mathcal{C}_5$    & 19.25   & 0.56 & 24.58 & --\\ \hline

$\mathcal{K}_{4}$  & 4.38   & 0.70 &  2.24 & --\\ \hline
Pattern $A$  & 61.36   & 11.85 &  52.77 &--\\ \hline
Pattern $B$  & 23.91   & 1.63 &  31.67 &--\\ \hline
Pattern $D$  & 481.25   & -- &  -- &--\\ \hline
 \end{tabular}
\end{table}

From \autoref{c_i} and \autoref{c_t}, 
we can see that RI algorithm outperforms all other measured algorithms. We can also say our algorithm is on par with LAD algorithm,
as the results of comparison of running times are similar, but vary instance from instance. 
Our algorithm nevertheless clearly outperforms another color coding algorithm, which on one hand solves more complicated problem
of weighted paths, but on the another, is still limited only to paths. 
Also, our algorithm is the only algorithm capable of enumerating disconnected patterns.

The weak point of the color coding approach (or possibly only of our implementation) appears to be the search for a pattern of larger size with very few (or possibly zero) occurrences. To achieve the desired error rate, we need to repeatedly run the algorithm many times. Therefore our algorithm takes longer time to run on some instances
(especially close to zero-occurrence ones), which are easily solved by the other algorithms. 

\subsection{ICPR2014 Contest Graphs}

To fully benchmark our algorithm without limitations on time or number of occurrences found, we perform a test on ICPR2014 Contest on Graph Matching Algorithms for Pattern Search in Biological Databases \cite{ICPR2014}.

In particular, we focus our attention on a \sc{Molecules} dataset, containing 10,000 (target) graphs representing the chemical structures of different small organic compounds and 
on a \sc{Proteins} dataset, which contains 300 (target) graphs representing the chemical structures of proteins and protein backbones.
Target graphs in both datasets are sparse and up to 99 vertices or up 10,081 vertices for \sc{Molecules} and  \sc{Proteins}, respectively.

In order to benchmark our algorithm without limiting its number of iterations, we focus on pattern graphs of small sizes, which offer reasonable number of iterations for an error rate of $\frac{1}{e}$.
Both datasets contain 10 patterns for each of considered sizes constructed by randomly choosing connected subgraphs of the target graphs.
We obtained an average matching time of all pattern graphs of a given size to all target graphs in a particular dataset.

\begin{table}[!h]
 \centering
 \caption[Comparison of average running time on ICPR2014 graphs]{Comparison of average running time on ICPR2014 graphs}
 \label{icpr_avg}
 \vspace{10px}
 \begin{tabular}{c|c||c|c|c}
Targets & Pattern	size & Our algorithm & LAD algorithm & RI algorithm\\
\hline \hline
\sc{Molecules} & 4 & 0.01 & 0.01 & 0.01\\ \hline
\sc{Molecules} & 8 & 0.67 & 0.14 & 0.01\\ \hline
\sc{Proteins} & 8 & 19.45 & 8.83 & 0.51\\ \hline
 \end{tabular}
\end{table}

From the results in Table \ref{icpr_avg}, we can see our algorithm being on par with LAD algorithm, while being outperformed by RI algorithm.
However, we mainly include these results as a proof of versatility of our algorithm.
As discussed in \cite{McCreeshPST18}, benchmarks created by constructing subgraphs of target graphs do not necessarily encompass the possible hardness of some instances
and might even present a distorted view on algorithms' general performance.
Thus, in the following benchmark we opt to theoretically analyze our algorithm. 

\subsection{Erd\H{o}s-R{\' e}nyi Graph Setup}

In order to precisely analyze the strong and weak points of our algorithm we measure its performance is a setting where both the pattern and the target are taken as an Erd\H{o}s-R{\' e}nyi random graph of fixed size with varying edge density and compare the performance of our algorithm with the analysis of McCreesh et al.~\cite{McCreeshPST18},
which focused on algorithms Glasgow, LAD, and VF2.

An Erd\H{o}s-R{\' e}nyi graph $G(n, p)$ is a random graph on $n$ vertices where
each edge is included in the graph independently at random with probability $p$.
We measure the performance on target graph of 150 vertices and pattern graph of 10 vertices with variable edge probabilities.
As our algorithm cannot be classified in terms of search nodes used (as in \cite{McCreeshPST18}), we measure the time needed to complete 10 iterations of our algorithm.

\begin{figure}[!h]
  \centering
  \includegraphics[width=0.6\textwidth]{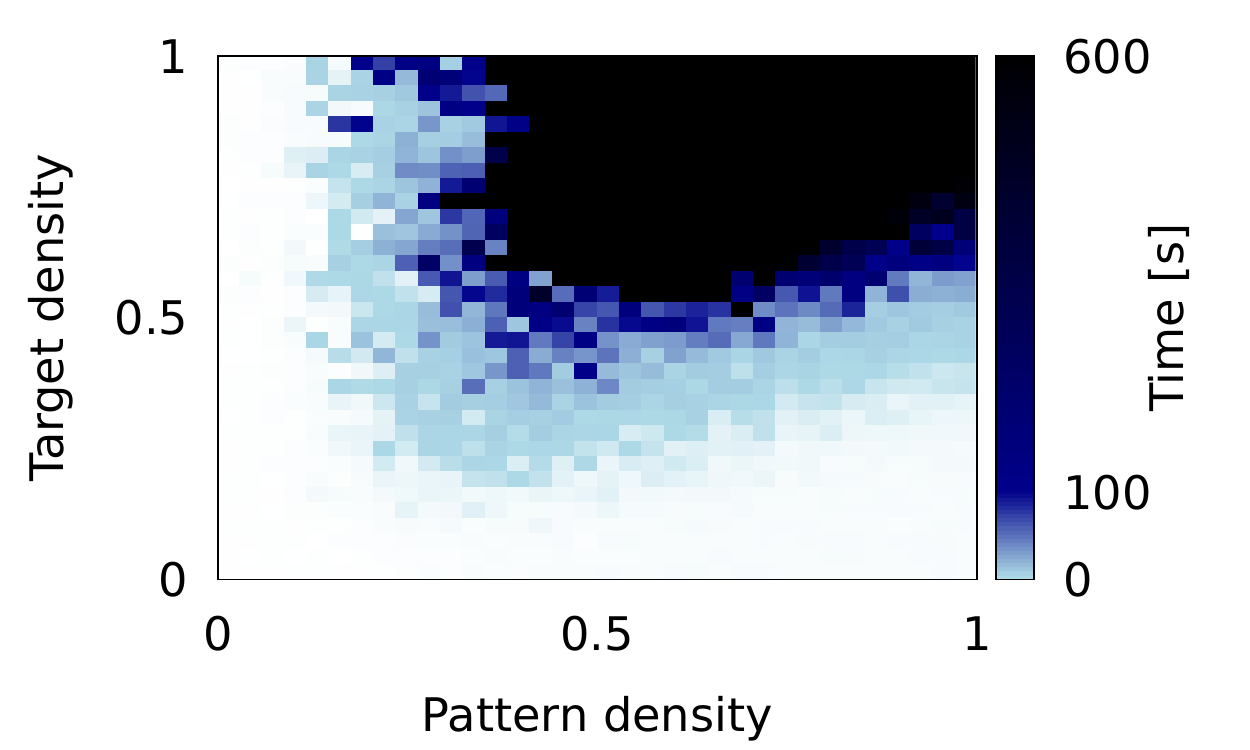}
  \caption{Behavior for target graph of 150 vertices and pattern graph of 10 vertices. 
           The x-axis is the pattern edge probability, the y-axis is the target edge probability, from 0 to 1 with step of 0.03.
           Graph shows the time required for our algorithm to complete 10 iterations (the darker, the more time is required).
           Black regions indicate instances on which a timeout of 600\,s occurred.}
  \label{rg10}         
\end{figure}

\begin{figure}[!h]
  \centering
  \includegraphics[width=0.6\textwidth]{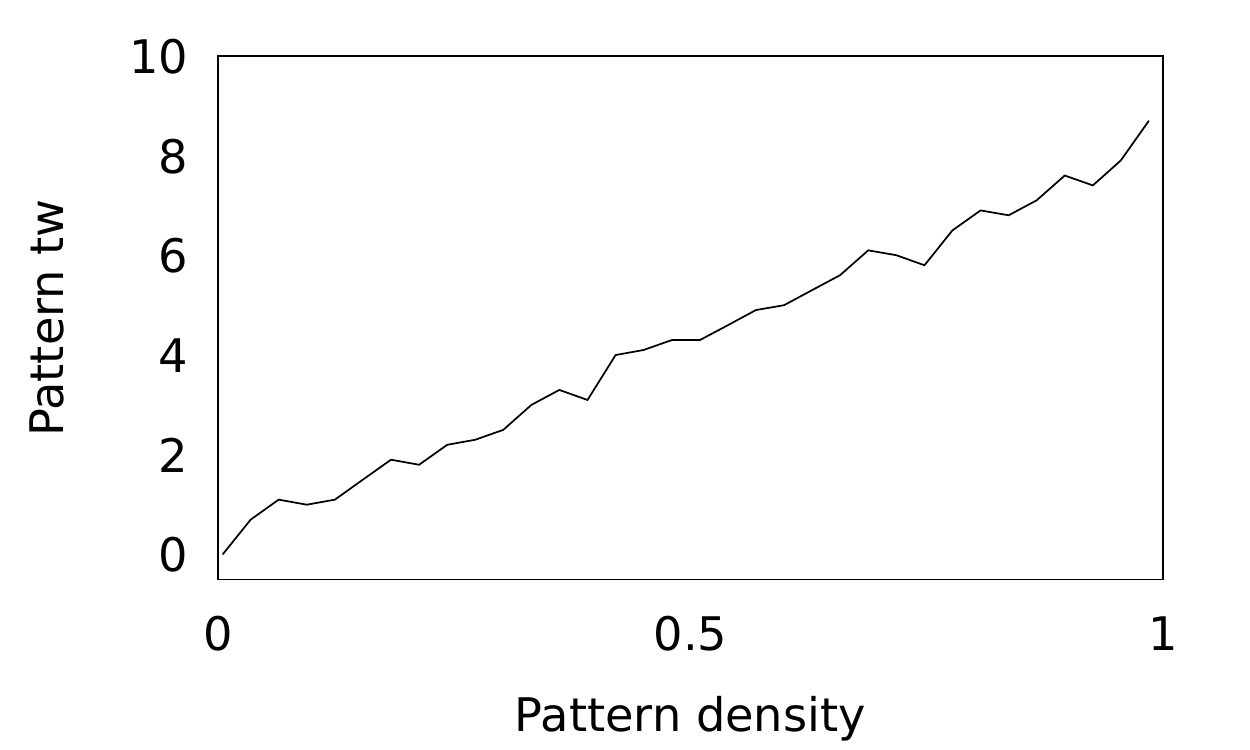}
  \caption{Correspondence of treewidth to the edge probability of a pattern graph with 10 vertices.}
  \label{tw10}
\end{figure}

\begin{figure}[!h]
  \centering
  \includegraphics[width=0.6\textwidth]{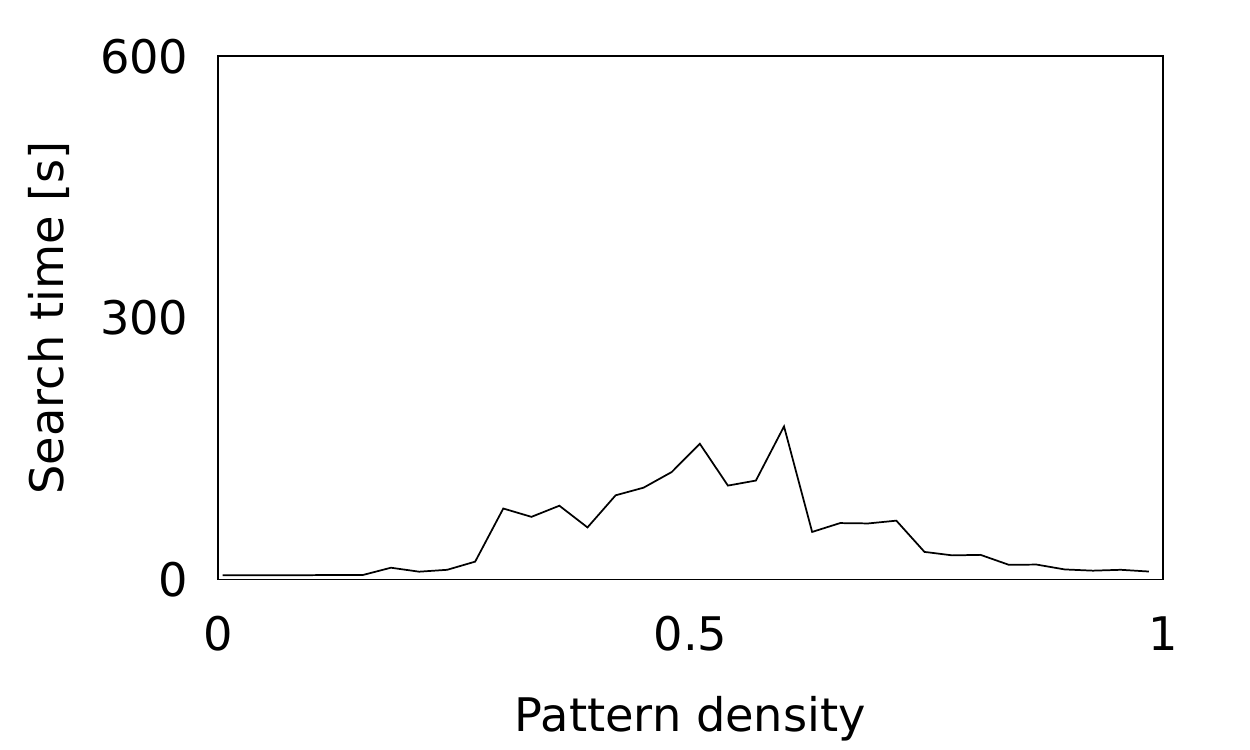}
  \caption{Time needed to complete 10 iterations of our algorithm on a target graph of 150 vertices with edge probability of 0.5 and pattern graph of 10 vertices
           with variable edge probability.}
  \label{msr_10_50}
\end{figure}

From Fig.~\ref{msr_10_50} we can see our algorithm indeed follows a well observed phase transition (transition between instances without occurrence of the pattern and with many occurrences of the pattern).
If we compare our results from Fig.~\ref{rg10} to the results of \cite{McCreeshPST18}, we can see that hard instances for our algorithm start
to occur later (in terms of edge probabilities).
However, due to the almost linear dependency of treewidth on edge probabilities (see Fig. \ref{tw10}),
hard instances for our algorithm concentrate in the ``upper right corner'' of the diagram, which contains dense graphs with naturally large treewidth.

\begin{figure}[!h]
  \centering
  \includegraphics[width=0.6\textwidth]{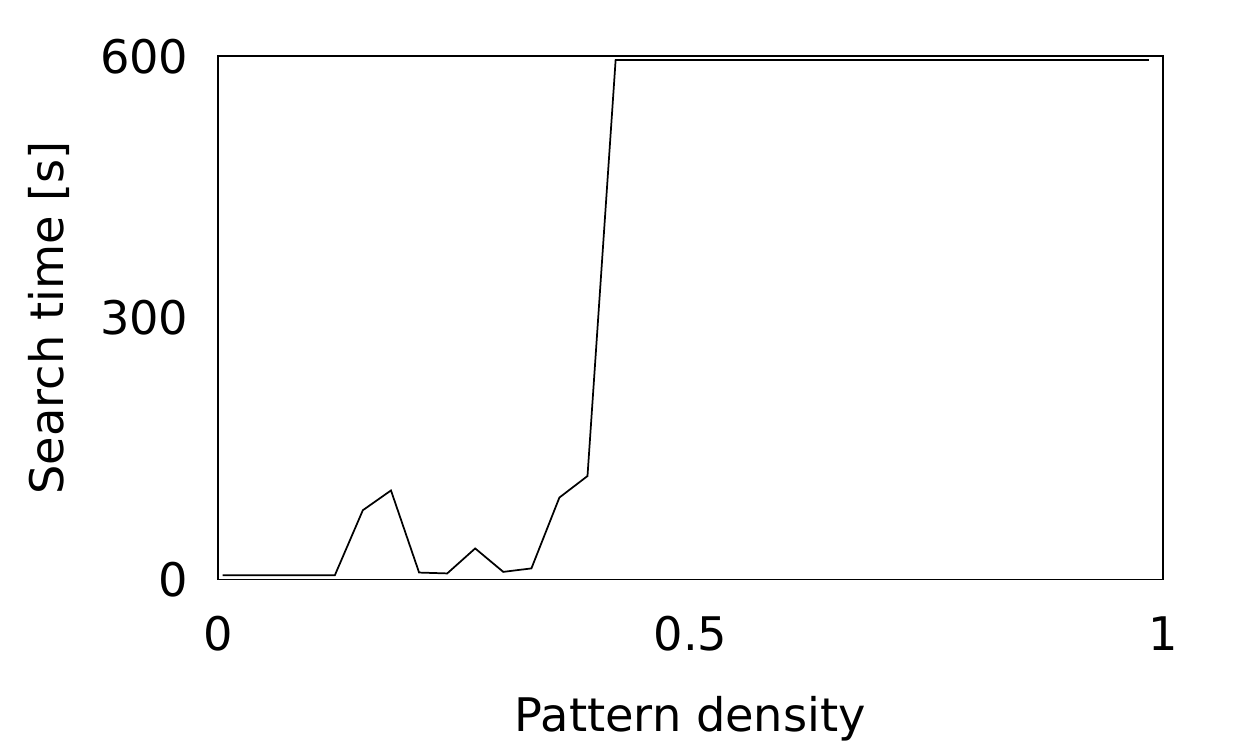} 
  \caption{Time needed to complete 10 iterations of our algorithm on a target graph of 150 vertices with edge probability of 0.8 and pattern graph of 10 vertices
           with variable edge probability.}
  \label{msr_10_80}         
\end{figure}

Therefore, it seems that our algorithm complements the portfolio of algorithms studied by Kotthoff et al.~\cite{KotthoffMS16} by an algorithm suitable just below the phase transition (in view of Fig.~\ref{rg10}).

\section{Conclusion}
\label{sec:conclusion}

We described an efficient implementation of the well known color coding algorithm for the subgraph isomorphism problem. 
Our implementation is the first color-coding based algorithm capable of enumerating all occurrences of patterns of treewidth larger than one. Moreover, we have shown that our implementation is competitive with existing state-of-the-art solutions in the setting of locating small pattern graphs. 
As it exhibits significantly different behaviour than other solutions, it can be an interesting contribution to the portfolio of known algorithms~\cite{KotthoffMS16,McCreeshPST18}.

As an obvious next step, the algorithm could be made to run in parallel. We also wonder whether the algorithm could be significantly optimized even further, possibly using some of the approaches based on constraint programming.

\bibliographystyle{splncs04}
\bibliography{main_arxiv}

\end{document}